\documentclass{sig-alternate-05-2015}
\usepackage{subfigure}

\begin{document}
\CopyrightYear{2016} 
\setcopyright{rightsretained} 
\conferenceinfo{WebSci '16}{May 22-25, 2016, Hannover, Germany} 
\isbn{978-1-4503-4208-7/16/05}
\doi{http://dx.doi.org/10.1145/2908131.2908185}

\renewcommand\ttlfnt{\fontfamily{phv}\fontseries{b}\fontsize{18}{22}\selectfont}

\title{C{\fontsize{17.28}{20}\selectfont\textsuperscript{3}}-index: Revisiting Authors' Performance Measure
%\titlenote{(Produces the permission block, and copyright information). For use with SIG-ALTERNATE.CLS. Supported by ACM.}
}
% \subtitle{[Short Paper]
% \titlenote{A full version of this paper is available as \textit{Author's Guide to Preparing ACM SIG Proceedings Using \LaTeX$2_\epsilon$\ and BibTeX} at \texttt{www.acm.org/eaddress.htm}}
% }

\numberofauthors{2} 
\author{
\alignauthor 
   Dinesh Pradhan, Partha Sarathi Paul, Umesh Maheswari, Subrata Nandi\\
   Department of CSE, NIT Durgapur, India\\
   \texttt{\{dineshkrp,mtc0113,umeshmaheswari7,subrata.nandi\}@gmail.com}
\alignauthor 
   Tanmoy Chakraborty\\
   University of Maryland, College Park, USA\\
  \texttt{tanchak@umiacs.umd.edu}
}

\CopyrightYear{2016} 
\setcopyright{rightsretained} 
\conferenceinfo{WebSci '16}{May 22-25, 2016, Hannover, Germany} 
\isbn{978-1-4503-4208-7/16/05}
\doi{http://dx.doi.org/10.1145/2908131.2908185}

\maketitle

\begin{abstract}
Author performance indices (such as h-index and its variants) fail to resolve ties while ranking authors with low index values (majority in number) which includes the young researchers. In this work we leverage the citations as well as collaboration profile of an author in a novel way using a weighted multi-layered network and propose a variant of page-rank algorithm to obtain a new author performance measure, $C^3$-index. Experiments on a massive publication dataset reveal several interesting characteristics of our metric: (i) we observe that $C^3$-index is consistent over time, (ii) $C^3$-index has high potential to break ties among low rank authors, (iii) $C^3$-index can effectively be used to predict future achievers at the early stage of their career.
\end{abstract}

\section{Introduction}
Easy access of publications via web has increased their visibility, leading the volume of authors and their publications increased exponentially, especially in computer science (CS) domain, during last decade~\cite{ChakrabortySTGM13} and made ranking authors harder. An effective index may help nominating an outstanding researcher for award, allocating research grants, etc. One may ask -- \emph{Is it possible to design an evaluation index for authors by combining multiple features, such as citation count, impact of coauthors, citing authors' profiles, etc such that not only high performers but also performance of the beginners (including young researchers) can be quantified unambiguously?} 

\begin{figure}[!t]
 \begin{minipage}[!t]{0.25\textwidth}
  \centering
  \subfigure[ Number of authors plotted against h-index for different years]{
    \includegraphics[scale=.33]{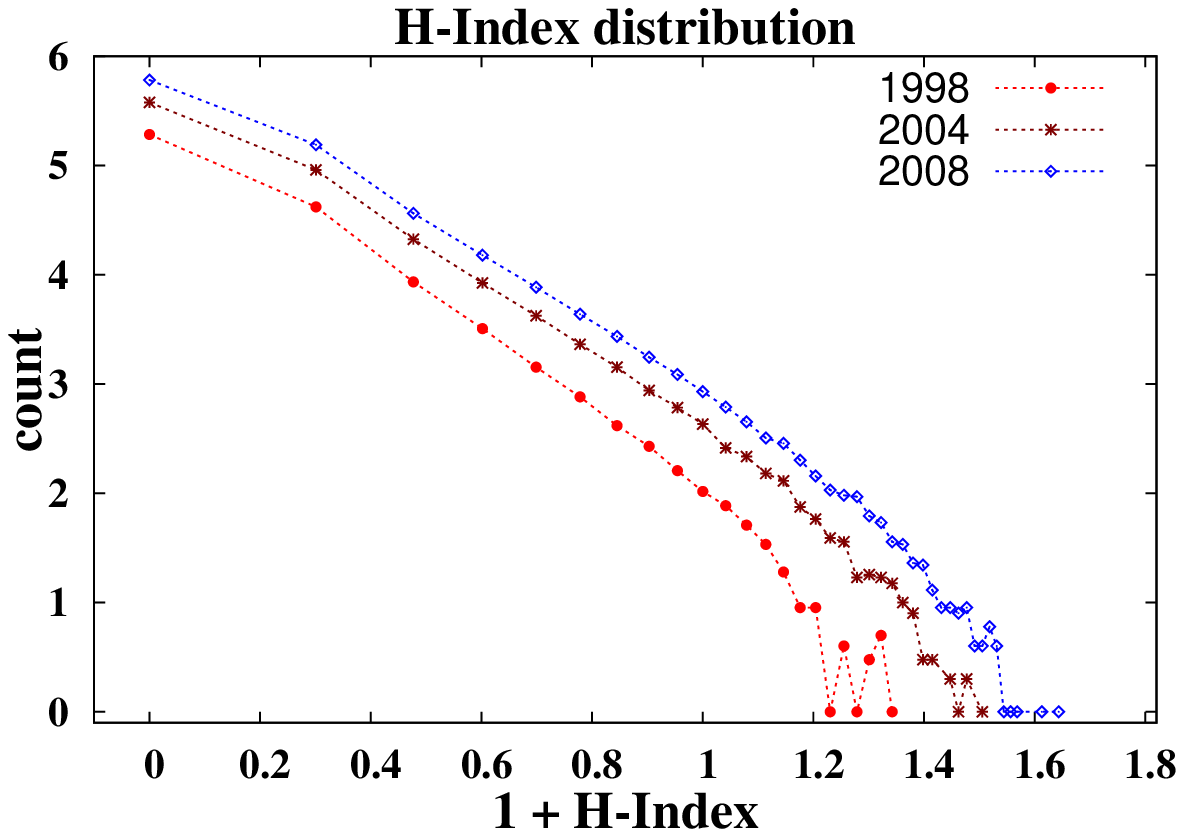}
    \label{subfigure:h_distribution}
    }
    \subfigure[\% of authors distributed over h-index bins in 1998 (blue-shaded bars); distribution of same pool in 2008 (red-shaded bars)]{
    \includegraphics[scale=.33]{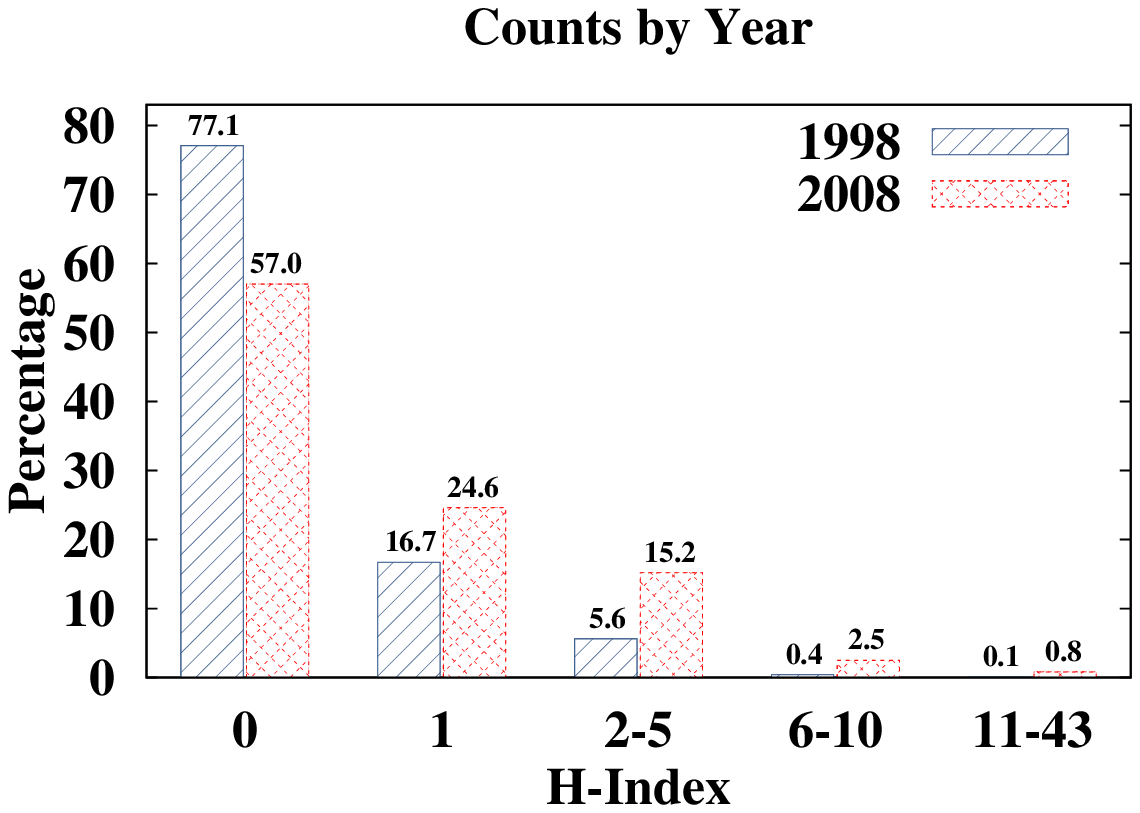}
    \label{subfigure:Bar_plot}
   }
 \end{minipage}
 \begin{minipage}[!t]{0.22\textwidth}
   \centering
    \subfigure[Network model used for $C^3$-index. The model comprises of three layers representing three relationships: author-author coauthorship (weighted, undirected), paper-paper citation (directed, unweighted) and author-author citation(weighted, directed). There are cross-layer directed edges from authors to papers.]{
    \includegraphics[scale=.058]{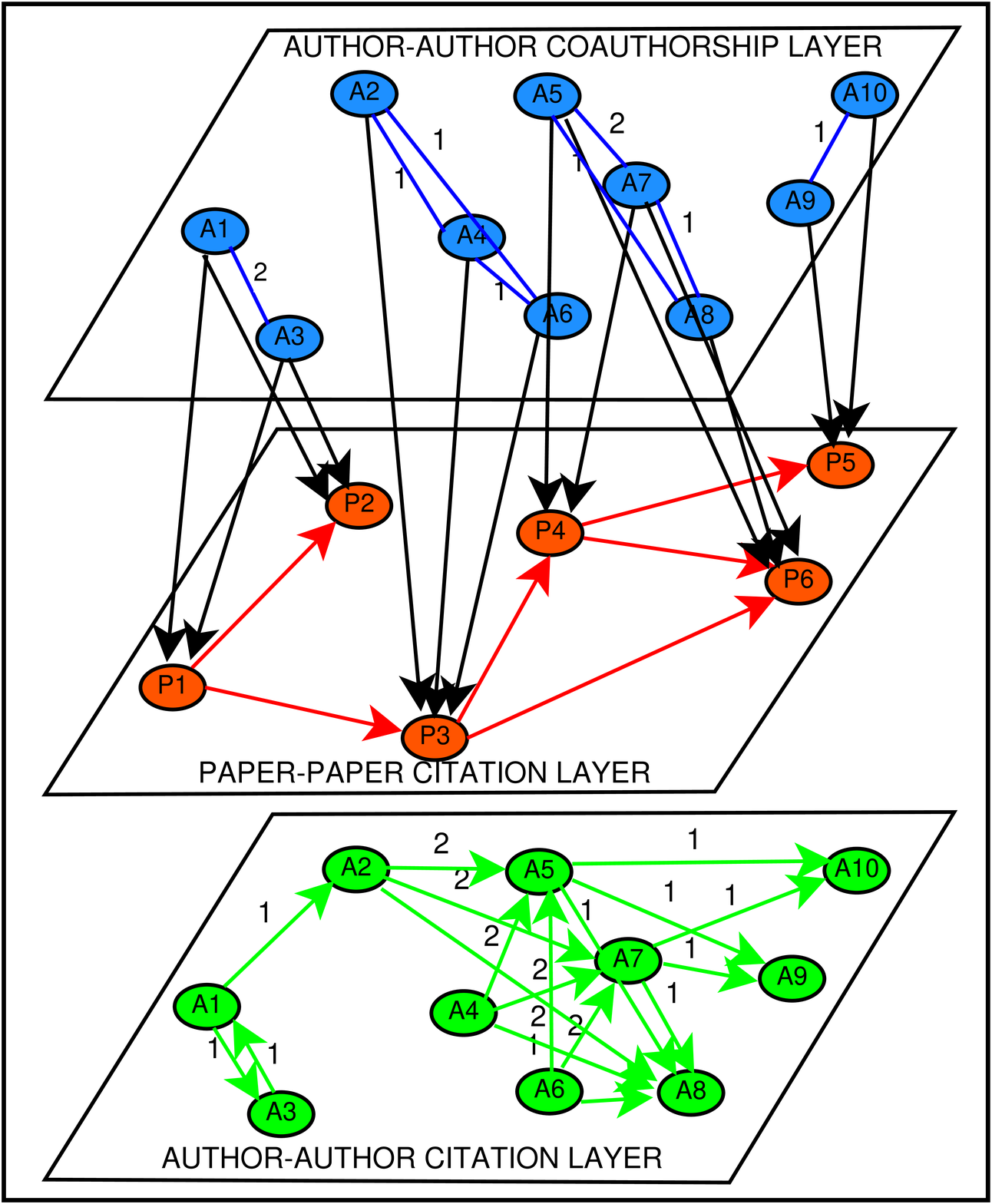}
    \label{subfigure:Network_Model}
    }    
 \end{minipage}
\caption{Evidences related to h-index and our proposed multi-layered model}
\label{figure:Evidence}
\vspace{-8mm}
\end{figure}

Most popular way of ranking authors is based on citations their works receive from peers. Based on citation count, several elegant yet simple indices exist: h-index, g-index, rank-citation index~\cite{h-index_variants,Petersen}, and so on. A closer look in Fig. \ref{subfigure:h_distribution} reveals a heavily skewed distribution of CS author count corresponding to the h-index they achieve as observed over a decade spanning from 1998 to 2008. In 2008, $\approx$80\% authors have h-index only upto 2; max-scale 43.  The drift of h-index of 1998 author set observed in 2008 is also very small (Fig. \ref{subfigure:Bar_plot}). The same holds true for other derivatives of h-index. Hence, such indices are unable to provide sufficient resolution to bulk authors (having low index) including young promising researchers who possibly have published few good papers and would receive enough citations afterwards~\cite{Chakraborty:2015,Chakraborty:2014}.

The limitations of citation count can be compensated by considering other features like (a) co-authors' profile - here an author's score is directly proportional to her co-authors' score, assuming that high-performers will publish with promising ones, (b) citing authors' profile - here an author's score is directly proportional to citing authors' score, assuming that high-profile authors refer to quality papers. In contrast to h-index and its derivatives, there are approaches which model above features as author-author collaboration and author-author citation networks and applied page-rank based algorithms~\cite{pagerank_variants_in_citation_networks} to rank authors. Obtaining a scoring function by modeling and combining multiple features is tricky; and unfortunately, multi-layered modeling has not been tried much. Moreover, existing literature broadly concerned on the dynamics of high and medium profile authors, and strangely ignored the bottom-liners who represent the bulk.

In this work we model the citations of papers and authors as well as coauthorship profile of authors in a novel way using a weighted multi-layered network and propose a variant of page-rank algorithm to rank authors. Our evaluation score is consistent over years and can effectively be used for ranking authors and predict early risers. In the next section we explain the dataset, the underlying network model and ranking algorithm followed by the results. 

\section{Experiment Setup and Results}

\noindent{\bf Dataset.} We use the dataset available in Arnetminer Project \cite{ChakrabortySTGM13}, containing 2,244,021 papers published between 1960 and 2013. After preprocessing, we consider 1,421,121 papers and 833,306 authors respectively in the years 1960-2008. While considering the impact of an author till year $T$ we only consider the evidences from our dataset till year $T$.

\noindent {\bf Network Model.} To utilize all the above features in parallel, we use network model shown in Fig \ref{subfigure:Network_Model}.

%we use three-layer network model referred in Fig. \ref{figure:Network_Model}. Layers are: author-author undirected, weighted coauthorship network (layer 1), paper-paper directed, unweighted citation network (layer 2) and author-author directed, weighted citation network (layer 3). The cross-layer arcs between layers 1 and 2 represent author-paper correspondence.

\noindent{\bf Measuring \emph{$C^3$-index}.} The strategy we propose is called \emph{$C^3$-index} (abbreviation of paper-paper {\bf C}itations, author-author {\bf C}itations and author-author {\bf C}ollaborations), where the author score $ C^{3}(t) $  at iteration $t$ is obtained as the normalized sum described by: $C_j^{3(t)} = (1-\theta) + \theta \times (ACI_j^{(t)}\ +\ AAI_j^{(t)}\ +\ PCI_j^{(t)})$. Here $ \theta $, set to 0.5 in our experiments, is the \textit{damping factor} for the page-rank based strategy. The component scores, {\bf A}uthor {\bf C}itation {\bf I}ndex (\textit{ACI}), {\bf A}uthor co{\bf A}uthorship {\bf I}ndex (\textit{AAI}) and {\bf P}aper {\bf C}itation {\bf I}ndex (\textit{PCI}) are obtained in iteration $t$ using:

\noindent $ACI_j^{(t)} = (1-\theta) + \theta \times \sum_{A_k\:\in\:C(A_j)}\dfrac{ACI_k^{(t-1)}}{outdeg(A_k)}$

\noindent $AAI_j^{(t)} = \sum_{A_k\:\in\:CA(A_j)} \dfrac{AAI_k^{(t-1)}}{deg(A_k)}$

\noindent $PCI_j^{(t)} = \biggl(C_j^{3(t-1)}\biggl)^{\alpha} \times \sum_{P_k\:\in\ P(A_j)\:} 												\dfrac{PQI_k^{(t-1)}}{\sum_{A_l\:\in\ A(P_k)\:} \biggl(C_l^{3(t-1)}\biggl)^{\alpha}} $, where \textit{PQI}, refers to as {\bf P}aper {\bf Q}uality {\bf I}ndex for a paper, is obtained as: $PQI_i^{(t)} = (1-\theta) + \theta \times \sum_{P_k\:\in\:C(P_i)} \dfrac{PQI_k^{(t-1)}}{outdeg(P_k)}$. Here, $ P(A_i) $ is set of papers of author $ A_i $, $ C(A_i) $ is set of authors citing author $ A_i $, and $ CA(A_i) $ is set of authors coauthoring with $ A_i $.

The parameter $ \alpha $ decides the way credit a paper being shared among coauthors. We set $ \alpha $ to 0, meaning that all the coauthors will receive equal share. But other values of $ \alpha $ may be tried, the credit then will be shared based on current $ C^3 $-Index of the coauthors.

\begin{figure}[!t]
 \begin{minipage}[!t]{0.25\textwidth}
  \centering
  \subfigure[ $C^3$-index vs. normalized h-index for 1998 author pool (Inset: same for 2008)]{
    \includegraphics[scale=.33]{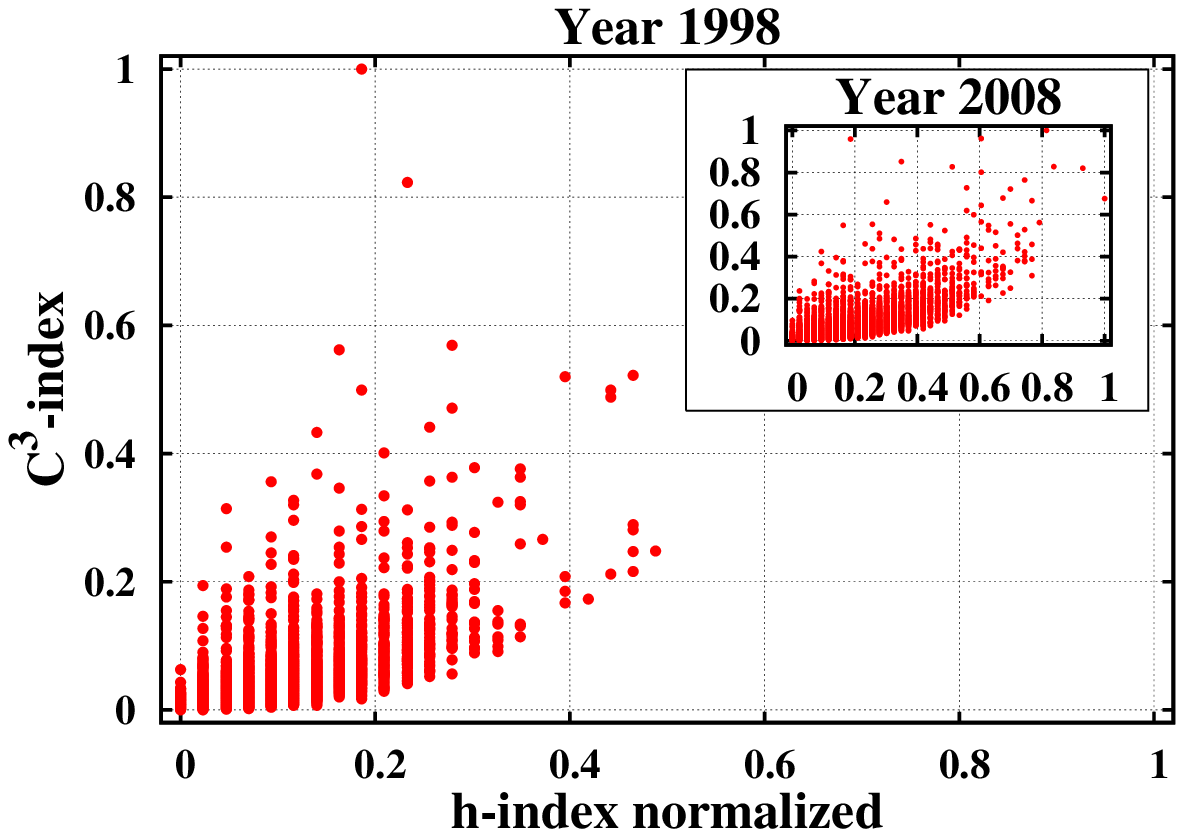}
    \label{subfigure:Scatter_Plot}
    }
 \end{minipage}
 \begin{minipage}[!t]{0.20\textwidth}
   \centering
    \subfigure[Growth of $C^3$-index against h-index from 1998 to 2008 of authors marked in Table \ref{Table:ECC_Component_Distribution}]{
    \includegraphics[scale=.33]{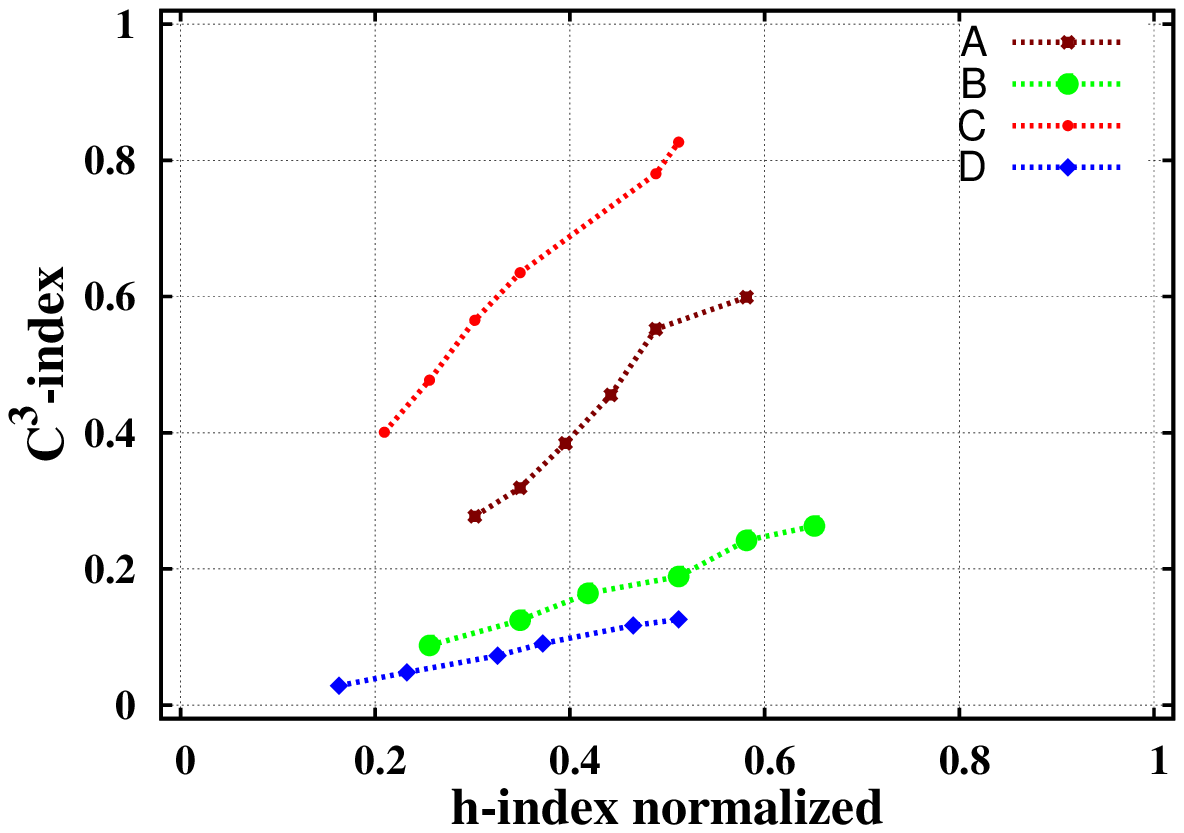}
    \label{subfigure:Author_Flight}
    }    
 \end{minipage}
\caption{Evidence related to existence of correlation of $C^3$-index with Growth of Authors over time}
\label{figure:Prediction}
\vspace{-5mm}
\end{figure}

\noindent{\bf Results.} Scatter-plots in Fig. \ref{subfigure:Scatter_Plot} show $C^3$-index for authors against their h-index for citation dataset relevant till 1998 (Inset: till 2008). We observe large pile of authors for each h-index values; however, one can notice that the authors bearing same h-index are sufficiently dispersed along the vertical scale. This indicate that $C^3$-index may be used to break the ties between authors having same h-index. Moreover, we observe large cluster of points close to the diagonal, indicating strong reasonable correlation among $ C^3$-index and h-index. 

\begin{table}[!h]
\tiny
\renewcommand{\arraystretch}{1.1}
\centering
\begin{tabular}{|c|c|c|}
\hline
Author & H-index & ACI, PCI, AAI\\
\hline
B. Bollobas & 1 & 0.45, 4.68, 2.54\\
\hline
\textbf{B Shneiderman (A)} & 13 & 23.12, 18.12, 13.42\\
\hline
G. Rozenberg & 4 & 2.94, 14.44, 9.21\\
\hline
\textbf{H. V. Jagadish (B)} & 11 & 6.50, 5.70, 5.24\\
\hline
M. S. Hsiao & 4 & 0.78, 0.64, 0.58\\
\hline
\textbf{Ronald L. Rivest (C)} & 9 & 39.58, 28.07, 11.17\\
\hline
S. Shelah & 2 & 0.44, 8.29, 6.24\\
\hline
\textbf{Tova Milo (D)} & 7 & 2.26, 1.74, 1.86\\
\hline
\end{tabular}
\caption{H-index and $C^3$-index component scores of eight authors from 1998 author pool}
\label{Table:ECC_Component_Distribution}
\vspace{-4mm}
\end{table}

To understand the inconsistent points of the diagram we choose eight points from Fig. \ref{subfigure:Scatter_Plot} and find their h-index and $ C^3$-index component scores (refer to Table \ref{Table:ECC_Component_Distribution}). We observe h-index having strong correlation with ACI scores, but weaker correlation with the others, inferring that h-index could capture ACI score well, but tends to ignore the others. These components are related to coauthorship layer, and may provide additional information that h-index is unable to furnish. Surprisingly, we further observe that the ranking of authors (especially those having very low h-index) (layer 1) based on $C^3$-index at early stage of their careers has high resemblance in later time periods. For instance, ranking of authors (with h-indices 1 and 2) in 1998 based on $C^3$-index has high correlation (Perason coefficient) with the same in 2004 (0.75, 0.74) and 2008 (0.55,0.50). Therefore, we anticipate that $C^3$-index can further be used to predict the successful authors as observed in Fig. \ref{subfigure:Author_Flight}, where growth of the same authors are shown till 2008.

%\begin{figure}[!h]
% \centering
% \graphicspath{{Figures/}}
%  \includegraphics[scale=.23]{Figure_Set_2}
% \caption{(a) $C^3$-index vs. normalized h-index for 1998 author pool (Inset: same for 2008), (b) Growth of $C^3$-index against h-index from 1998 to 2008 of authors marked in Table \ref{Table:ECC_Component_Distribution}}
% \label{figure:Figure_Set_2}
%\vspace{-5mm}
%\end{figure}

\section{Conclusion}
The proposed multi-feature based evaluation score successfully resolves ambiguity among the major class of low profile authors which we believe is a major contribution. The impact of growth/failure/saturation of an author score needs to be studied more systematically with respect to features like topical influence in which the author primarily works. A thorough investigation is required to show the universality of our findings for other domains. One can think of categorizing the authors based on their future prospect.

% \bibliographystyle{abbrv}
% \bibliography{sigproc}  % sigproc.bib is the name of the Bibliography in this case

\end{document}